# Plane Wave Density Functional Theory Studies of the Structural and the Electronic Properties of Amino Acids Attached to Graphene Oxide via Peptide Bonding

Byeong June Min[†], Chang-Woo Lee[§], and Hae Kyung Jeong[†]

[†]*Department of Physics,* [§]*Department of Biomedical Science,*

*Daegu University, Kyungsan 712-714, Republic of Korea*

We studied the electronic and the structural properties of amino acids (alanine, glycine, and histidine) attached to graphene oxide (GO) by peptide bonding, via plane wave pseudopotential total-energy calculations within the local spin density approximation (LSDA). The HOMO-LUMO gap, the Hirshfeld charges, and the equilibrium geometrical structures exhibit keen variations depending on the species of the attached amino acid. The GO-amino acid system appears to be a good candidate as a biosensor.

# I. INTRODUCTION

Immobilization of protein on a graphene system is an exciting research topic that leads to many important applications in biomedical technology [1-4]. This is due to the unique electronic properties of graphene as a two-dimensional zero gap material. Its electronic structure exhibits linear dispersion near the Fermi level, making the system extremely sensitive to a charge transfer. Since the electronic structure of a graphene-amino acid system is likely to change sensitively as an amino acid is attached to graphene, we may hope to fabricate a sensitive biosensor that can be used in the diagnosis and treatment of many diseases.

However, immobilization of protein on pure graphene is a challenge because of the chemical inertness of graphene. The interaction between the protein and the graphene is mostly due to $\pi-\pi$ interaction [5]. Since a stronger bonding is preferable for the fabrication of a robust biosensor, protein attached to graphene oxide (GO) via peptide bonding has attracted interest [6].

GO may be considered as graphene decorated with functional groups, such as hydroxyl, epoxy, and carboxyl group. The carboxyl group of GO is protonated at low pH, but, as the pH increases, the carboxyl group dissociates yielding a carboxylate ion. It is possible to form a peptide bond between the carboxyl group of GO and the amino group of amino acid, releasing a water molecule. Peptide bonds are higher in energy but very stable, such that the spontaneous hydrolysis takes an extremely long time.

We intend to investigate the structural and the electronic properties of three selected amino acids, alanine (Ala), glycine (Gly), and histidine (His), covalently attached to GO, via plane wave density functional theory calculations. Since amino acids are the building blocks of protein, we hope to provide theoretical insights that may be applicable for the much bigger problem of GO-protein systems. Besides, the GO-amino acid system in itself is an interesting topic from the basic science point of view and may even have its own application.

## II. CALCULATION

We constructed a cluster model of GO by assembling 36 carbon atoms in a graphene-like configuration and passivating the dangling bonds with 16 hydrogen atoms, and then adding an epoxy group and a hydroxyl group. The presence of an epoxy group and a hydroxyl group introduces a conspicuous curvature in the GO cluster, implying a fair amount of elastic energy. A carboxyl group is added at the edge of the GO cluster, and then, an amino acid (Ala, Gly, or His) is connected to the cluster via peptide bonding.

The initial configuration is built using the Avogadro package [7]. The functional groups and amino acids are attached to the GO cluster group by group, its geometry being optimized under the universal force field (UFF) [8] at every stage as each functional group is added to the cluster. The GO cluster reacts quite sensitively as another group of atoms are attached, as can be expected from the linear dispersion of the electronic structure of graphene near the Fermi level.

The resulting configuration is used as the starting position in the *ab initio* molecular dynamics simulation, performed using the ABINIT package [9]. A periodic boundary condition is imposed assuming a cubic box of 18.52 Å for each dimension, with gamma point sampling. Norm-conserving pseudopotentials of Troullier-Martins type [10] with Teter parameterization of exchange-correlation functional [11] were used within the local spin density approximation (LSDA). Plane wave energy cutoff of $980\,eV$ is used. Self-consistency cycles are repeated until the difference of the total energy becomes smaller than $2.72 \times 10^{-6}\,eV$, twice in a row. The system is relaxed until the average force on the atoms become smaller than $2.57 \times 10^{-3}\,eV/\text{Å}$.

## III. RESULTS AND DISCUSSION

The present GO model that has an epoxy group, a hydroxyl group, and a carboxyl group is shown, together with GO-alanine (GO-Ala), GO-glycine (GO-Gly), and GO-histidine (GO-His) structure determined from the present calculation (Fig. 1). We find a big change in the bond length of the carboxyl group in the amino acids (shown in black in Fig. 1) as the amino acid is attached to the GO. The bond length changes to 1.682 Å for GO-Ala, 1.625 Å for GO-Gly, and 1.598 Å for GO-His, compared to 1.514 Å for Ala, 1.498 Å for Gly, and 1.512 Å for His. The bond length increase is 0.168 Å for Ala, 0.127 Å for Gly, and 0.086 Å for His. If we consider the bond length of the COO group in free standing amino acids (with the H atom removed from the carboxyl), we observe an even more distinguished difference. The bond lengths in this case are 1.531 Å in Ala, in 1.527 Å Gly, and 1.729 Å in His. Relative to the deprotonated amino acids, the bond length change upon attachment to GO would be 0.151 Å for Ala, 0.098 Å for Gly, and -0.131 Å for His. The results are summarized in Table 1. Despite the similarity between the amino acids, their structures change in a drastically different manner when the H atom is removed from them and, more importantly, when they are attached to GO. Side views of GO-amino acid systems are shown in Fig. 2. It appears that Ala orients more in the perpendicular direction to the GO plane than the others.

Hirshfeld net charge analysis also shows distinct characteristics. The COO group in GO-Ala has a charge of -0.38e, compared to -0.42e in GO-Gly and -0.45e in GO-His. To verify the amount of the charge transfer due to the attachment to GO, we calculated the Hirshfeld net charges of the free-standing amino acids with the H atom removed from the carboxyl group. It turns out that the COO group in Ala has a charge of -0.28e, Gly -0.28e, and His -0.29e. Then, taking these values as the reference, the additional charge transfers are estimated as -0.10e in GO-Ala, -0.14e in GO-Gly, and -0.16e in GO-His.

The total charge transfer from the amino acids to the GO varies characteristically, too. The Hirshfeld net charges of the carbon atoms in the GO sheet, hydroxyl group, epoxy group, and the amino acids are summarized in Table 2 and Fig. 3. The Hirshfeld net charges of the amino acids were -0.507e for GO-His, -0.387e for GO-Gly, and -0.356e for GO-Ala. The carbon atoms in the GO sheet had net charges of -0.461e (GO-Ala), -0.437e (GO-Gly), and -0.348e (GO-His), compared to -0.690e of carbon atoms in the original GO model. The net charge of the epoxy group was almost without variation. In contrast, the charge of the hydroxyl group had a smaller magnitude but showed a larger variation. The net charges of the passivating H atoms are 0.971e (GO-Ala), 0.980e (GO-Gly), and 0.988e (GO-His), compared to 0.886e (GO model). The H atoms should represent the response from the larger GO sheet that cannot be treated explicitly within our present model system.

The HOMO-LUMO gap of GO-Ala, GO-Gly, and GO-His are 0.533 eV, 0.496 eV, and 0.129 eV, respectively, compared to 4.221 eV, 4.275 eV, and 3.563 eV of Ala, Gly, and His alone (Table 3). Again, the HOMO-LUMO gaps of the free-standing amino acids become more distinct as they are attached to GO. Especially, GO-His has a very small HOMO-LUMO gap. The density of states profile is shown in Fig. 4.

The energetics of GO-amino acid systems also depend sensitively on the species of the amino acids. The binding energy, as defined by

$E_b$ = E(GO´+Amino Acid) - E(GO´) - E(Amino Acid) + E($H_2O$),

where GO´ denotes GO without the H atom in the carboxyl group, varies significantly. $E_b$ was 0.32 eV for GO-Ala, 0.10 eV for GO-Gly, and -0.02 eV for GO-His. These are quite large deviations from the typical peptide bond energy 0.16 eV. Our results suggest that GO-amino acid bonds are not the ordinary peptide bonds.

It is quite a pleasant surprise that albeit all the outward appearances of similarity between these amino acids, their equilibrium structure and electronic properties show keen distinction. Our results are encouraging for the prospect of building a GO based biosensor system.

## IV. CONCLUSION

We studied the structural and electronic properties of amino acids, alanine, glycine, and histidine, attached to GO via plane wave pseudopotential calculations. The similarity of the equilibrium structures and the electronic properties of the amino acids disappear as they are attached to GO. The equilibrium structure and the electronic properties change in a quite distinct manner, depending on the species of the amino acids. Such sensitivity would be a welcome feature for the application in the field of biosensing.

## ACKNOWLEDGMENTS

B. J. Min is much indebted to the librarians of Daegu University and wishes to express special thanks for their great service. B. J. Min acknowledges financial support from the Daegu University Research Funds.

Table 1. The bond lengths (in Å) of the carboxyl group of GO-Ala, GO-Gly, and GO-His, compared to the those of protonated and deprotonated amino acids.

|     | GO-attached | protonated amino acid | deprotonated amino acid |
|-----|-------------|-----------------------|-------------------------|
| Ala | 1.682       | 1.514                 | 1.531                   |
| Gly | 1.625       | 1.498                 | 1.527                   |
| His | 1.598       | 1.512                 | 1.729                   |

Table 2. The Hirshfeld net charges (in units of e) of the amino acids, the carbon atoms, the epoxy group, and the hydroxyl group of GO.

|        | Amino acid | Carbon atoms | Epoxy group | Hydroxyl group |
|--------|------------|--------------|-------------|----------------|
| GO+Ala | -0.356     | -0.461       | -0.121      | -0.032         |
| GO+Gly | -0.387     | -0.437       | -0.120      | -0.035         |
| GO+His | -0.507     | -0.348       | -0.111      | -0.021         |

Table 3. The HOMO-LUMO gap (in eV) of Ala, Gly, and His attached to the GO compared with those of standalone Ala, Gly, and His.

|     | standalone | GO-attached |
|-----|------------|-------------|
| Ala | 4.221      | 0.533       |
| Gly | 4.275      | 0.496       |
| His | 3.563      | 0.129       |

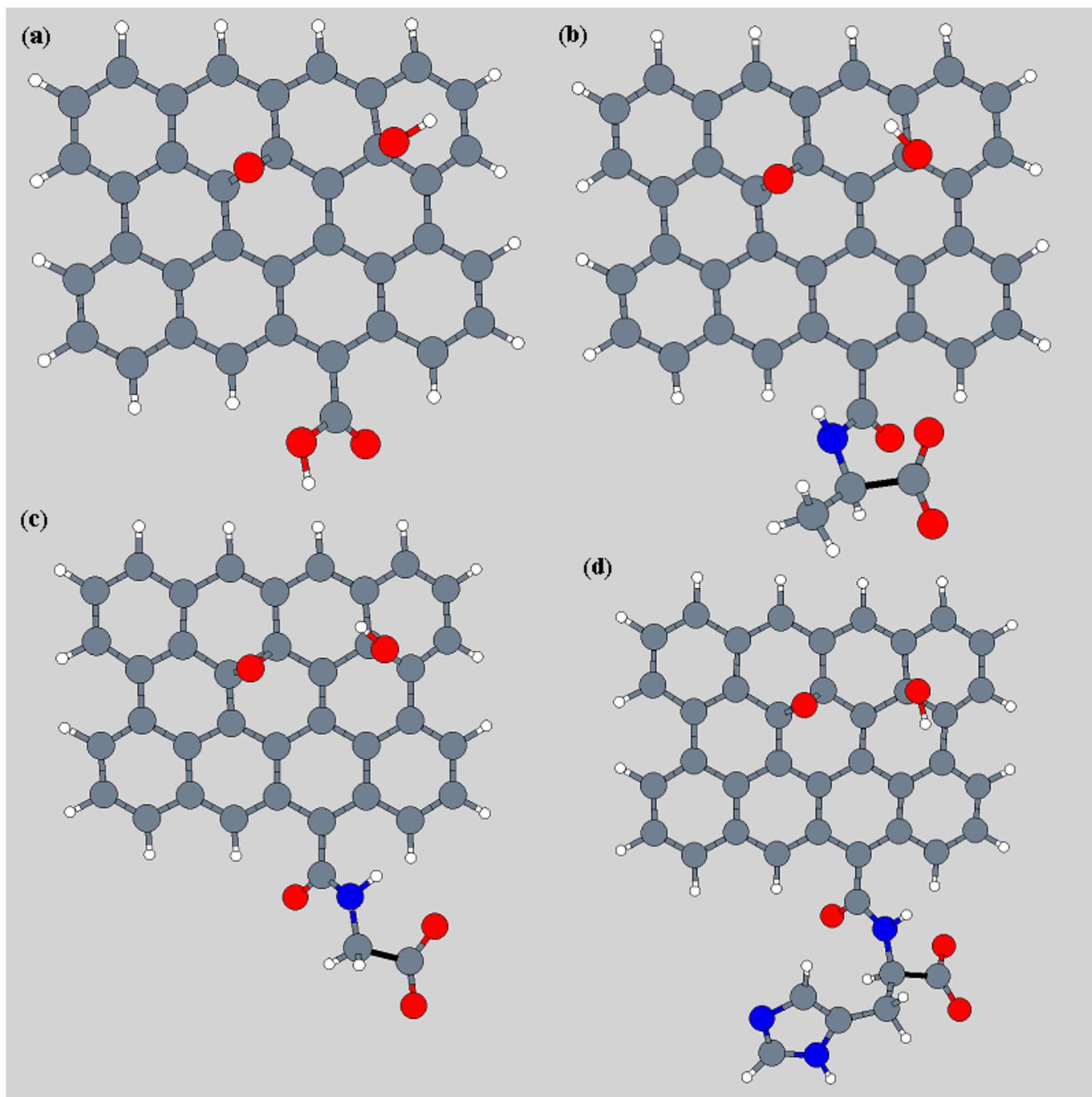

Fig. 1

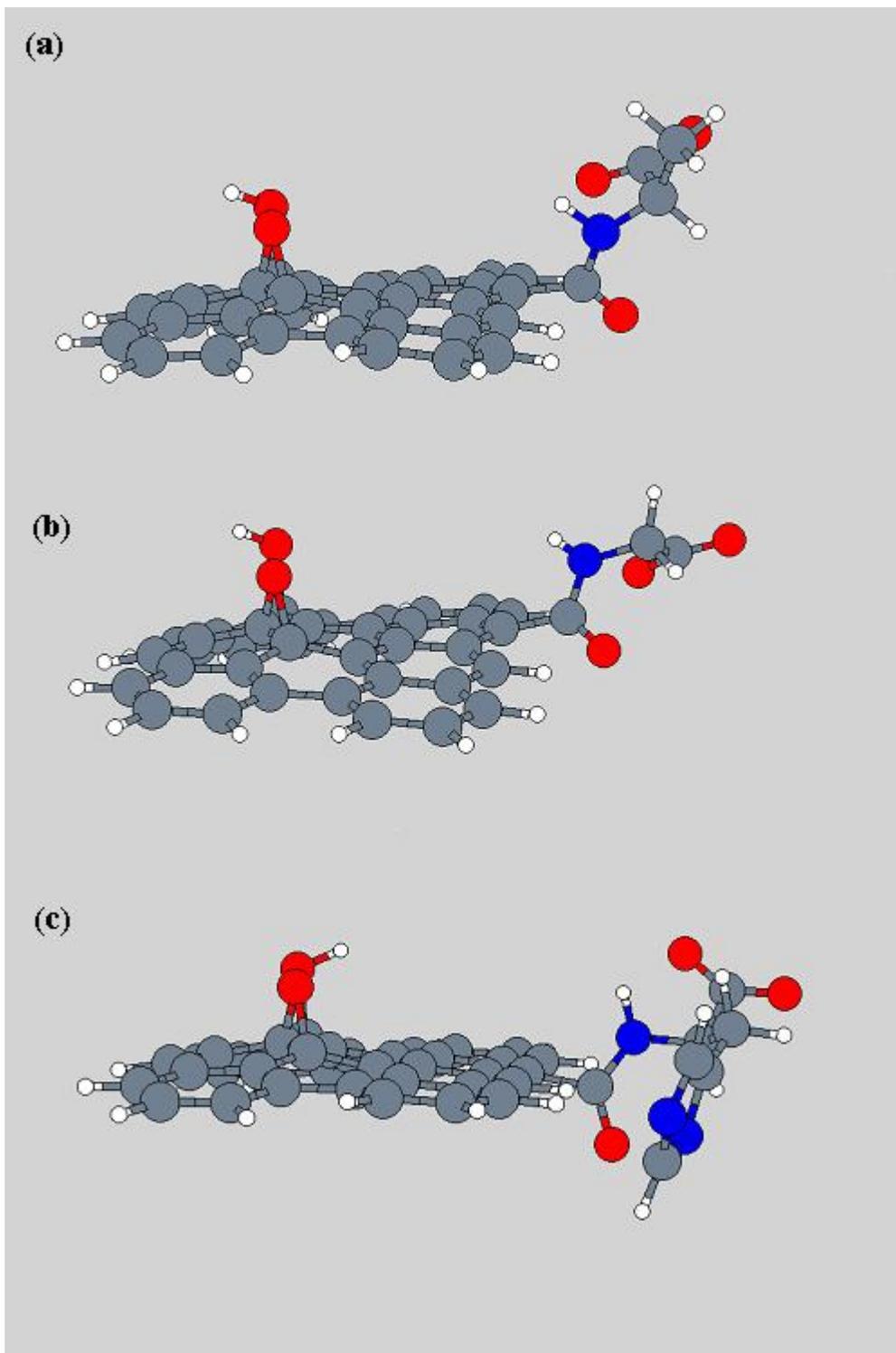

Fig. 2

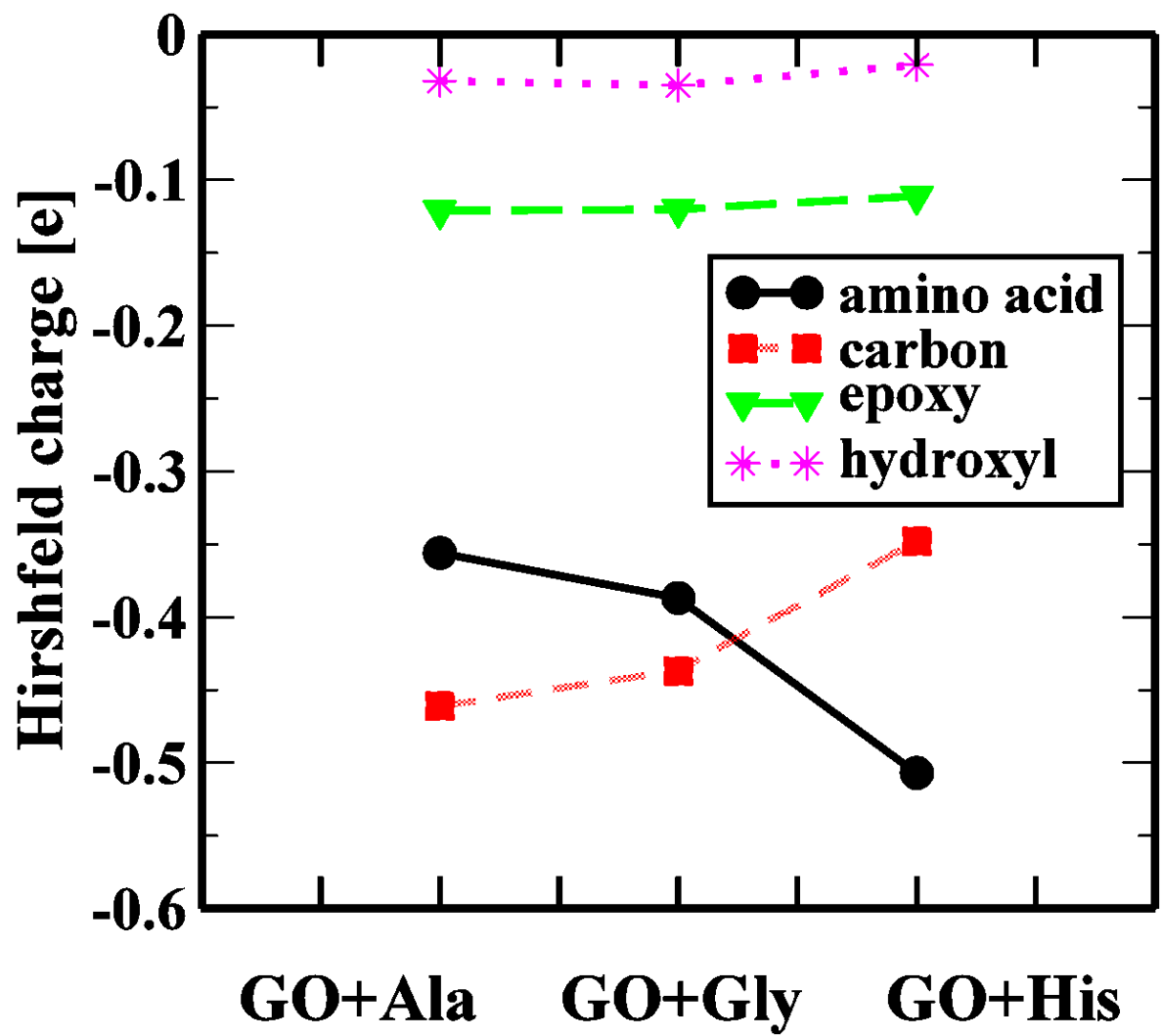

Fig.3

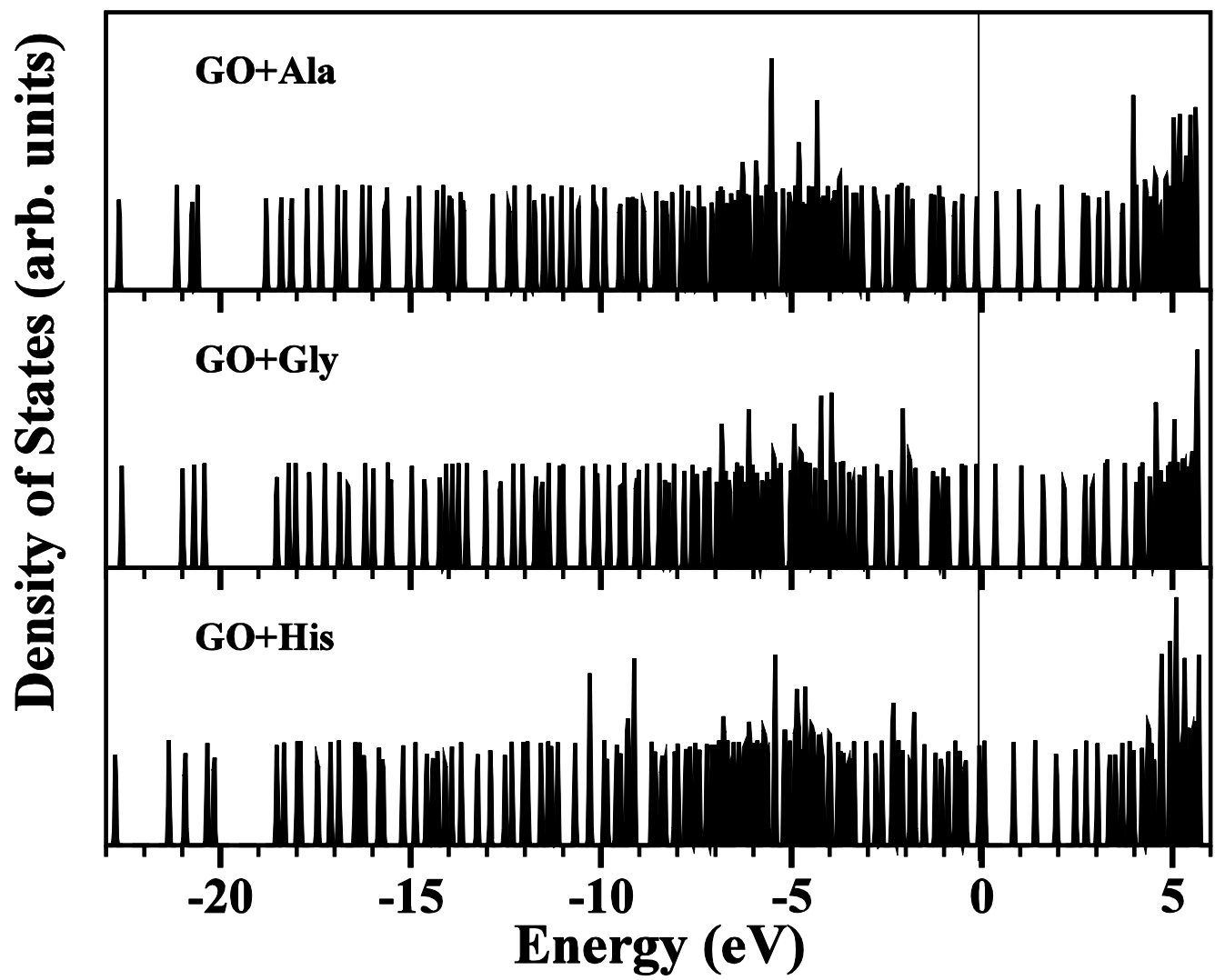

Fig. 4

Figure Captions.

Fig. 1. The equilibrium geometries of (a) GO, and GO with (b) Ala, (c) Gly, and (d) His. The bond that elongates when the amino acid attaches to GO is shown in black.

Fig. 2. Side views of (a) GO-Ala, (b) GO-Gly, and (c) GO-His.

Fig. 3. The Hirshfeld net charges of the carbon atoms in the graphene sheet, hydroxyl group, epoxy group, and the amino acids for GO-Ala, GO-Gly, and GO-His.

Fig. 4. The density of states profile for GO-Ala, GO-Gly, and GO-His. The HOMO-LUMO gap of GO-His is much smaller than those of GO-Ala and GO-Gly.